\documentclass[prb,twocolumn,showpacs,preprintnumbers,amsmath,amssymb]{revtex4}

\usepackage{graphicx}
\usepackage{dcolumn}
\usepackage{bm}

\begin{document}

\title{Higher-order mesoscopic fluctuations in quantum wires:\\
Conductance and current cumulants}%
\author{Markku P. V. Stenberg}
\email{markku.stenberg@tkk.fi}
\author{Jani S\"arkk\"a}%
\affiliation{%
Laboratory of Physics, Helsinki University of Technology, P.O. Box 4100,
FIN-02015 HUT, Finland
}%
\date{\today}
\begin{abstract}
We study conductance cumulants $\langle \! \langle g^n\rangle\!\rangle$ 
and current cumulants $C_j$ related to heat and electrical transport in 
coherent mesoscopic quantum wires near the diffusive regime. We consider the
asymptotic behavior in the limit where the number of channels and the length 
of the wire in the units of the mean free path are large but the bare 
conductance is fixed. A recursion equation unifying the descriptions of the 
standard and Bogoliubov--de Gennes (BdG) symmetry classes is presented. We 
give values and 
come up with a novel scaling form for the higher-order conductance cumulants. 
In the BdG wires, in the presence of time-reversal symmetry, for the cumulants 
higher than the second it is found that there may be only contributions which 
depend nonanalytically on the wire length. This indicates that diagrammatic 
or semiclassical pictures do not adequately describe higher-order spectral 
correlations. Moreover, we obtain the weak-localization corrections to $C_j$ 
with $j\le 10$.
\end{abstract}
\maketitle

\section{\label{sec:introduction} Introduction}
In a quantum wire (quasi-one-dimensional geometry), the quantities of
interest, such as conductance, are not selfaveraging. 
\cite{dittrich97} 
Thus, a statistical description of a variety of quantum effects, such as
universal conductance fluctuations, requires the conductance cumulants
$\langle\!\langle g^n \rangle\!\rangle$ or conductance distribution $P(g)$
to be studied. Cumulants of current $C_j$, 
like $C_2$ that is directly related to shot noise, 
also yield information which is not contained in the 
averaged conductance. Higher-order mesoscopic fluctuations, though smaller, 
provide important additional information about the transport process.
For example, the recent experiments \cite{mohanty02} where an anomalously 
asymmetric $P(g)$ was detected have spurred dispute \cite{deba} and
interest on higher conductance cumulants in mesoscopic wires. This is due to 
the fact that noninteracting one-parameter scaling models suggest that near 
the diffusive region the conductance distribution $P(g)$ is close, but not 
identical, to Gaussian shape. \cite{altshuler86} For the moment, 
current-cumulant measurements on the third cumulant $C_3$ have been 
carried out \cite{reulet03} and a scheme for a detection of $C_4$ has recently 
been put forward. \cite{ankerhold05}

The fundamental symmetries of the Hamiltonian of a disordered conductor
manifest themselves in the statistical properties of the energy levels
and particle states. The implications of symmetry are conveniently
analyzed within the framework of random matrix theory. \cite{rmt}
There are altogether ten different random matrix theories 
for different symmetric spaces \cite{caselle04}
of the Hamiltonian. \cite{altland97}
The symmetry classes are customarily referred by the Cartan's symbol
for the symmetric space of the corresponding Hamiltonian. \cite{altland97} 
In this paper we consider seven symmetry classes: A(I,II), C(I), and D(III).
Disordered normal metals may be studied through
standard or Wigner-Dyson (WD) universality classes A(I,II).
\cite{mehta91} The BdG classes C(I) and D(III) are appropriate, e.g., for 
``disorder-facilitated'' quasiparticle transport \cite{senthil98} in 
unconventional superconductors. Compared to the WD 
ensembles, an extra degree of freedom arises for the BdG classes, since a 
distinction between 
particlelike and holelike quasiparticles of the BdG formalism is made.
Further, the systems are classified according to the presence or absence of 
time-reversal (TR) and spin-rotation (SR) symmetry. The corresponding 
symmetric spaces are characterized by the multiplicities of the ordinary and 
long roots \cite{helgason78} denoted $m_0$ and $m_l$ 
(see, e.g., Table 1 of Ref.~\onlinecite{titov01}). 
In addition to the two symmetry parameters, the degeneracy $d$ of 
transmission and reflection eigenvalues has to be taken into account. 
Besides having definite values for certain symmetries,
$m_0$ and $m_l$ may also be considered as interpolation parameters 
between different symmetry classes. In the absence of electron-electron 
interactions, phase randomization and local maximum entropy principles imply
in a quasi-one-dimensional geometry a scaling equation, the
Dorokhov-Mello-Pereyra-Kumar (DMPK) equation. \cite{dmpk} 
The DMPK equation yields the evolution of $P(g)$ as a function of dimensionless
length $s$. We set $s=L/N\ell$, where $L$ is the length 
of the wire, $N$ is the number of channels, and $\ell$ is the elastic mean free 
path. In a wire geometry, the DMPK equation has also been generalized for 
the BdG structures \cite{brouwer00} and for the so-called chiral
classes. \cite{brouwer98}

Near metallic regime the most important mechanisms inducing deviations
from Gaussian $P(g)$ are weak localization (WL) and weak antilocalization 
(WAL). WL and WAL result from the interference of the closed trajectories of 
the electron and their time-reversed conjugates.
Constructive interference (WL) leads to a decrease of $g$
while destructive interference (WAL) enhances conductance.
In a conductor with length shorter or comparable with
the coherence length, in the absence 
of magnetic field, destructive interference 
may be realized in a material with strong spin-orbit scattering whereas 
spin-rotation symmetry leads to constructive interference. The W(A)L 
corrections to conductivity have been used to study experimentally the 
quantum coherence in mesoscopic structures 
(see, e.g., Ref.~\onlinecite{trionfi04} 
for recent measurements). It is very difficult to evaluate the exact
influence of W(A)L on the values of $\langle\!\langle g^n \rangle\!\rangle$ 
with $n>3$ by such field theoretic approaches as diagrammatic techniques
\cite{altshuler86, rossum97} or by the nonlinear sigma model. \cite{efetov97}

A nonperturbative treatment of the first and second cumulants based on 
Fourier analysis on the supersymmetric manifold was given in 
Ref.~\onlinecite{mirlin94} for WD classes. The behavior of the lowest 
conductance 
cumulants $\langle\!\langle g^{1,2} \rangle\!\rangle$ has been intensively 
computed for the WD ensembles ($m_l=1$) recently, especially in the 
metal-insulator crossover region 
(see, e.g., Refs.~\onlinecite{perez02} and \onlinecite{muttalib03}). 
In Ref.~\onlinecite{altshuler86}, based on a $2+\epsilon$ expansion,
the expression 
\begin{equation}
\label{eq:abrik}
\langle\!\langle g^n \rangle\!\rangle\sim \langle g\rangle^{2-n}, \qquad n<1/s
\end{equation}
was presented for the standard unitary ensemble \mbox{($m_0=1$, $m_l=1$)}.
Likewise, for the symmetry class AI, but for a quasi-one-dimensional 
conductor (where one has \mbox{$\langle g\rangle\approx 1/s$}), the same 
equation was put forward in Ref.~\onlinecite{tartakovski95}. This formula has 
been widely accepted until now.

For the BdG wires near metallic region, the dependence of $\langle g \rangle$
on $m_0$ and $m_l$ may be found in Ref.~\onlinecite{brouwer00}. 
Leading terms of $\langle\!\langle g^2 \rangle\!\rangle$
(universal conductance fluctuations) and some correction terms for 
$\langle\!\langle g^2 \rangle\!\rangle$ and for $C_2$ 
may be found in Ref.~\onlinecite{imamura01}. An essential singularity has 
been shown to occur for the first and second cumulants in the small $s$ 
expansion in the chiral unitary class. \cite{mudry99}
Mac\^edo considered in Ref.~\onlinecite{macedo02} the cases with $m_0=2$ and 
$0\le m_l\le 2$ and found that the third cumulant contains only a component 
which is nonanalytic in $s$ at $s=0$. Localization in superconducting wires 
has been discussed in Refs.~\onlinecite{brouwer00} and
\onlinecite{gruzberg05}. The conductance cumulants with $n\le 4$ were 
computed in Ref.~\onlinecite{perez02} for the WD
classes A and AI by a Monte Carlo method.

In this paper we present a recursion equation unifying the description
of the WD and BdG symmetry classes and yielding the cumulants of the order 
$n<1/s$. For the two BdG classes with TR symmetry we find that the 
higher-order cumulants ($n\ge3$) contain no contributions that are analytic 
in $s$ at $s=0$. 
We elucidate the dependence of the
higher cumulants on the universality class and 
give values for $\langle\langle g^n\rangle\rangle$ in terms of $m_0$ and 
$m_l$. We emphasize that even though Eq.~(\ref{eq:abrik}) is correct 
for $n=1,2$, for \mbox{$2<n<1/s$} there exists a more appropriate expression, 
our Eq.~(\ref{eq:sc}).
Furthermore, we calculate the weak-localization corrections 
to the current cumulants $C_j$ with $j\le 10$. We consider a disordered 
quantum wire, i.e., a quasi-one-dimensional
geometry. For the BdG universality classes we study heat transport whereas for 
the WD classes our results apply also for electrical transport. We calculate 
the cumulants at zero temperature, zero frequency, and at low voltage
in the limit \mbox{$N\to\infty$},\mbox{$\ L/\ell\to\infty$},
\mbox{$s={\rm constant}$}, near 
the diffusive region, where one has $1/N\ll s\ll 1$.
\section{Cumulants}
\subsection{Method}
The starting point for our analysis is the generalized DMPK equation which 
reads \cite{dmpk, brouwer00} 
\begin{eqnarray}
\label{eq:dmpk}
\lefteqn{\partial_{s}w_{s}(\mbox{\boldmath{$\lambda$}})=
\frac{2N}{m_0 N+1+m_l-m_0}
\sum_{i=1}^{N}\frac{\partial}{\partial \lambda_i}
\bigg\{ }\nonumber\\
&&
\times 
[\lambda_i(1+\lambda_i)]^{(m_l+1)/2} J_{m_0}
(\mbox{\boldmath{$\lambda$}})
\frac{\partial}{\partial\lambda_i}
[\lambda_i(1+\lambda_i)]^{(1-m_l)/2}
\frac{w_s(\mbox{\boldmath{$\lambda$}})}{J_{m_0}(\mbox{\boldmath{$\lambda$}})}
\bigg\}.
\nonumber\\
\end{eqnarray}
The variables
$\{\lambda_i\}_{i=1}^{N}$ are related to the transmission 
probabilities $\{\tau_i\}_{i=1}^N$ of the channels $\{i\}_{i=1}^{N}$ 
through \mbox{$\tau_i=(1+\lambda_i)^{-1}$} while
$w_{s}(\mbox{\boldmath{$\lambda$}})$ is the distribution function
for \mbox{$\mbox{\boldmath{$\lambda$}}=(\lambda_1,\ldots,\lambda_N)$}.
The Jacobian $J_{m_0}(\mbox{\boldmath{$\lambda$}})$ is given by
\mbox{$J_{m_0}(\mbox{\boldmath{$\lambda$}})
=\prod_{i<j}|\lambda_i-\lambda_j|^{m_0}$}.
In a long wire, for the symmetry classes D(III), additional terms enter the 
DMPK equation. \cite{gruzberg05} Near diffusive regime such components, 
however, are irrelevant and we will ignore them. 

In order to calculate the cumulants we adopt a method reminiscent to
the moment expansion method introduced in Refs.~\onlinecite{mello88}
and \onlinecite{mello91}.
It is convenient to introduce the moment 
generating function $Z_s(\mbox{\boldmath{$\mathrm{q}$}})$ 
and the cumulant generating function (CGF)
$\varphi_{s}(\mbox{\boldmath{$\mathrm{q}$}})$,
\begin{eqnarray}
\label{eq:defcumul}
\ln Z_s(\mbox{\boldmath{$\mathrm{q}$}})&=&
\ln\langle\exp(-\mbox{\boldmath{$\mathrm{q}$}}
\cdot\mbox{\boldmath{$\mathrm{T}$}})\rangle_s=
\varphi_{s}(\mbox{\boldmath{$\mathrm{q}$}})
\end{eqnarray}
in order to systematically solve the DMPK equation.
Here one has $\mbox{\boldmath{$\mathrm{T}$}}=
(T_1,\ldots,T_N)$, $T_k=\sum_{i}^{N}\tau_{i}^{k}$. 
The expectation value $\langle\cdots\rangle_s$ is taken with
respect to the distribution of $\tau_i$s. The conductance cumulants may be
obtained from
\begin{equation}
\langle\!\langle g^n\rangle\!\rangle\equiv
(-d)^{n}\frac{\partial^{n} \varphi_{s}(\mbox{\boldmath{$\mathrm{q}$}})}{\partial q_1^{n}}
\bigg{|}_{\mbox{\boldmath{$\mathrm{q}$}=\boldmath{$0$}}}.
\label{ccdef}
\end{equation}
For large $N$ it is natural to seek the CGF in the form of the
expansion \cite{gopar95}
\begin{equation}
\varphi_{s}(\mbox{\boldmath{$\mathrm{q}$}})=
\sum_{j=-\infty}^{1}\varphi_{s}^{(j)}(\mbox{\boldmath{$\mathrm{q}$}})N^j
\label{nexp}.
\end{equation}
In Ref.~\onlinecite{gopar95} it was found that in such an expansion, 
for the symmetry class A, $\varphi_{s}^{(j)}(\mbox{\boldmath{$\mathrm{q}$}})$
is an odd (even) polynomial of degree $2-j$ in $\mbox{\boldmath{$\mathrm{q}$}}$
with $j$ odd (even).
For all the WD and BdG classes, in the region where 
$\varphi_{s}^{(j)}(\boldsymbol{\mathrm{q}})$ may be expanded in non-negative
powers of $L/\ell$, 
it may be shown by induction from 
Eq.~(\ref{eq:dsvarphi}) in Appendix A and from the definition of the CGF, 
Eq.~(\ref{eq:defcumul}),
that $\varphi_{s}^{(j)}(\boldsymbol{\mathrm{q}})$ is of the form
\begin{equation}
\varphi_{s}^{(j)}(\boldsymbol{\mathrm{q}}) =
\sum_{n=1}^{2-j}
\sum_{k_{1},\dots,k_{n}=1}^{\infty}
A_{k_{1},\dots,k_{n}}^{(j)}(L/\ell) \prod_{i=1}^{n}q_{k_{i}}.
\label{q2m}
\end{equation}
We seek $A_{k_{1},\dots,k_{n}}^{(j)}(L/\ell)$ in the form of a rational 
function of $L/\ell$. By using Eqs.~(\ref{eq:defcumul}) and (\ref{eq:dsvarphi})
it can be shown by induction that the further condition $1/N\ll s$ 
implies that $A_{k_{1},\dots,k_{n}}^{(j)}(L/\ell)$ takes the form
\begin{equation}
A_{k_{1},\dots,k_{n}}^{(j)}(L/\ell)=
a_{k_{1},\dots,k_{n}}^{(j)}(L/\ell)^{-j}+
\mathcal{O}\left[(L/\ell)^{-j-1}\right].
\end{equation}
Thus in the limit $N\to\infty,\ L/\ell\to\infty,\ s={\rm constant}$
we obtain for all the BdG and WD classes the expansion
\begin{eqnarray}
\label{eq:finalexp}
\varphi_{s}(\mbox{\boldmath{$\mathrm{q}$}})=
\sum_{j=-\infty}^{1}\sum_{n=1}^{2-j}
\sum_{k_1,\ldots,k_{n}=1}^{\infty}a_{k_1,\dots,k_{n}}^{(j)}s^{-j}
\prod_{i}^{n} q_{k_{i}}.
\end{eqnarray}
This is essentially an expansion in the inverse powers of large bare
conductance and it is expected to converge in the metallic regime
$1/N\ll s\ll 1$.
We will show that, in the presence of the TR symmetry,
$a_{k_1, \dots, k_{n}}^{(j)}$ with \mbox{$n>2$} actually vanish for
the BdG classes CI and DIII. Thus for the classes CI and DIII, the cumulants 
higher than the second contain no components that are analytic in $s$ at 
$s=0$. 

From Eq.~(\ref{eq:dsvarphi}) in Appendix A one obtains the first of the 
coefficients $a_{1}^{(1)}=-1$. For the other coefficients we obtain
after a lengthy calculation an algebraic recursive
equation (see Appendix A for more details),

\begin{widetext}
\begin{eqnarray}
\label{eq:recua}
\lefteqn{\bigg(-m+2\sum_{p=1}^{n} k_{p}\bigg)a_{k_1,\dots,k_n}^{(m)}=
\bigg[\sum_{j=m}^{1}
\sum_{n_{1}=\text{max}(1,n+m-j)}^{\text{min}(2-j,n)}
\sum_{p_1\neq p_2\neq\cdots\neq p_n=1}^{n}
k_{p_{1}}n_1(n-n_1+1)\frac{1}{n!}}
\nonumber\\
&&\times\bigg(
\sum_{l=0}^{k_{p_{1}}-1}
\Delta_{n_{1},n,l,k_{p_{1}}}^{j,m}
a_{k_{p_1}-l,k_{p_2},\ldots,k_{p_{n_{1}}}}^{(j)}
a_{l+1,k_{p_{n_{1}+1}},\ldots,k_{p_n}}^{(m-j+1)}
-\sum_{l=0}^{k_{p_{1}}-2}
a_{k_{p_1}-l-1,k_{p_2},\ldots,k_{p_{n_{1}}}}^{(j)}
a_{l+1,k_{p_{n_{1}+1}},\ldots,k_{p_n}}^{(m-j+1)}\bigg)\bigg]\nonumber\\&&
+\sum_{p=1}^{n}k_{p}\left\{
\left(\theta_1(n+1)\sum_{l=0}^{k_{p}-1}
a_{k_{p}-l,l+1,k_1,\ldots,k_{p-1},k_{p+1},\ldots,k_{n}}^{(m+1)}
+\frac{2\theta_2}{m_0 n}\sum_{r=1,r\neq p}^{n}
k_r a_{k_{p}+k_r+1,k_1,\ldots,k_{p-1},k_{p+1},\ldots,
k_{r-1},k_{r+1},\ldots,k_{n}}^{(m+1)}\right)\right.\nonumber\\
&& -
\left(k_{p}\rightarrow k_{p}-1 \right)
+\frac{\theta_3}{m_0}[(m_0-2)k_p+m_l-1]
a_{k_{p}+1,k_1,\ldots,k_{p-1},k_{p+1},\ldots,k_n}^{(m+1)}
-\frac{\theta_3}{m_0}[(m_0-2)k_p+2m_l-m_0]a_{k_1,\ldots,k_n}^{(m+1)}{\Bigg\}}.
\end{eqnarray}
\end{widetext}

Here we have 
\begin{equation*}
\begin{split}
a_{k_{p_1}-l,k_{p_2},\ldots,k_{p_{n_{1}}}}^{(j)}&=a_{k_{p_1}-l}^{(j)}
\quad{\rm for}\quad n_1=1,\\
a_{l+1,k_{p_{n_{1}+1}},\ldots,k_{p_n}}^{(m-j+1)}&=a_{l+1}^{(m-j+1)}
\quad{\rm for}\quad n_1=n.
\end{split}
\end{equation*}
Moreover, we denote 
\begin{equation*}
\begin{split}
&\Delta_{n_{1},n,l,k_{p_{1}}}^{j,m}=
(1-\delta_{j,m}\delta_{n_{1},n}\delta_{l,0})
(1-\delta_{j,1}\delta_{n_{1},1}\delta_{l,k_{p_{1}}-1}),\\
&\theta_1=\theta(-n-m),\quad
\theta_2=\theta(n-2)\theta(2-n-m),\\
&\theta_3=\theta(1-n-m)
\end{split}
\end{equation*}
with 
\begin{eqnarray*}
\theta(n) = \left\{
\begin{array}{ll}
1 &{\rm for}\quad n \ge 0,\\
0 &{\rm otherwise}
\end{array}
\right..
\end{eqnarray*}
Note also that $\sum_{l=0}^{k_{p_1}-2}\equiv 0$ with $k_{p_1}=1$.
For the higher cumulants this recursion relation is conveniently 
evaluated by a computer.
Because of Eq.~(\ref{eq:finalexp}), the 
coefficients $a_{k_1,\ldots,k_n}^{(m)}$ are 
invariant under the change of subindices. The coefficients are evaluated 
(1) in the  order of decreasing $m=1,0,-1,\ldots$, (2) for a given $m$ in 
the order of increasing $n$, and (3) for given $m$ and $n$ they may be 
calculated, e.g., in the order of increasing subindices.

Equation (\ref{ccdef}) implies
\begin{equation}
\langle\!\langle g^n\rangle\!\rangle
=(-d)^n n!\sum_{m=-\infty}^{1} a_{\underbrace{1,\dots,1}_{n}}^{(m)}
s^{-m}.
\label{ga}
\end{equation} 
It then follows that, instead of
\begin{equation}
\label{eq:scnaive}
\langle\!\langle g^{n} \rangle\!\rangle \sim s^{n+\delta_{m_0,2}-2},\qquad
2 < n < 1/s,
\end{equation}
that generalizes Eq.~(\ref{eq:abrik}) to the remaining six symmetry classes,
we obtain the scaling  
\begin{equation}
\label{eq:sc}
\langle\!\langle g^{n} \rangle\!\rangle \sim s^{n+\delta_{m_0,2}-1},\qquad
2 < n < 1/s,
\end{equation}
that is down by one power of $s$ relative to the result of Altshuler,
Kravtsov, and Lerner.

Calculations based on a nonperturbative microscopic approach 
\cite{tartakovski95} suggest that there exist additional nonperturbative 
contributions that become significant for \mbox{$n>1/s$} and which lead to 
log-normal tails of the conductance distribution but are not describable 
within the one-parameter scaling approach. Since we cannot produce 
such nonperturbative components by our method we restrict our considerations 
to conductance cumulants of the order \mbox{$n<1/s$}.

References \onlinecite{altshuler86} and \onlinecite{tartakovski95}
missed the cancellation of the terms of the order \mbox{$\mathcal{O}(s^{n-2})$}
since the numerical prefactors corresponding to $a_{k_1,\ldots,k_n}^{(2-n)}$ 
were not evaluated. The cancellation of the leading contribution of 
Eq.~(\ref{eq:abrik}) has already been pointed out for $n=3$ in 
Refs.~\onlinecite{gopar95}, \onlinecite{macedo94}, and \onlinecite{rossum97}. 
This cancellation is specific to quasi-one-dimensional geometry. 
\cite{rossum97} For higher $n$ the cancellation of the leading terms 
has been proved in Appendix B.
\subsection{Conductance cumulants}
For the numerical values of 
$\langle\!\langle g^n\rangle\!\rangle$, the cases $n=1,2$ have been covered, e.g., 
in Ref.~\onlinecite{macedo94} for the WD classes and in 
Ref.~\onlinecite{brouwer00} ($n=1$) and Ref.~\onlinecite{imamura01} ($n=1,2$) 
for the BdG classes. 
Our Eq.~(\ref{eq:recua}) provides an efficient derivation 
of these results. As an application of Eq.~(\ref{eq:recua}) we give the 
results for $n\le 6$,
\begin{widetext}
\begin{equation}
\begin{split}
\label{eq:gcumulants}
d^{-1}\langle\!\langle g \rangle\!\rangle =&
\frac{1}{s}+\frac{(m_{0}-2m_{l})}{3m_{0}}+[(3m_{0}-8m_{l})
(m_{0}-2)+4m_{l}(m_{l}-2)]\frac{s}{45m_{0}^2}\\
&+\{[-21m_{0}+6m_{0}^2+2m_{l}(-5m_{0}-4m_{l}+16)]
(m_{0}-2)+8m_{l}(m_{l}-1)(m_{l}-2)\}\frac{2s^2}{945m_{0}^3}+
\mathcal{O}(s^3),\\
d^{-2}\langle\!\langle g^{2} \rangle\!\rangle =&
\frac{2}{15m_{0}}+\frac{8(m_{0}-2)s}{315m_{0}^2}-
[(15+19m_{0}-69m_{l})(m_{0}-2)+32m_{l}(m_{l}-2)]
\frac{4s^2}{4725m_{0}^3}+\mathcal{O}(s^3),\\
d^{-3}\langle\!\langle g^3 \rangle\!\rangle=& 
-\frac{8(m_0-2)s^2}{1485m_0^{3}}
+[(172\ 635+200\ 793m_0-751\ 117m_l)(m_0-2)+353\ 792m_l(m_l-2)]
\frac{32s^3}{638\ 512\ 875m_0^4}\\
&+\{[-291\ 960-1\ 371\ 774m_0+554\ 337m_0^2+(2\ 303\ 952-
1\ 429\ 012m_0+205\ 356m_l)m_l](m_0-2)\\
&+286\ 720m_l(m_l-1)(m_l-2)\}\frac{8s^{4}}{212\ 837\ 625m_0^5}
+\mathcal{O}(s^5),\\
d^{-4}\langle\!\langle g^4\rangle\!\rangle=& 
\frac{512(m_0-2)s^{3}}{155\ 925m_0^{4}}
-[(17\ 006\ 315+15\ 370\ 851m_0-62\ 563\ 249m_l)(m_0-2)+
29\ 630\ 464m_l(m_l-2)]\\
&\times\frac{32s^4}{54\ 273\ 594\ 375m_0^5}+\mathcal{O}(s^5),\\
d^{-5}\langle\!\langle g^5\rangle\!\rangle=& 
-\frac{11\ 229\ 952(m_0-2)s^{4}}{3\ 919\ 486\ 725m_0^{5}}+
[(55\ 989\ 824\ 345+45\ 233\ 187\ 091m_0-192\ 229\ 424\ 511m_l)(m_0-2)\\
&+91\ 546\ 451\ 968m_l(m_l-2)]
\frac{128s^{5}}{510\ 443\ 155\ 096\ 875m_0^6}+\mathcal{O}(s^6),\\
d^{-6}\langle\!\langle g^6\rangle\!\rangle=&
\frac{6\ 045\ 601\ 792(m_0-2)s^{5}}{1\ 747\ 609\ 738\ 875m_0^{6}}
+\mathcal{O}(s^6).
\end{split}
\end{equation}
\end{widetext}
For a normal-metal wire ($m_l=1$), the term of the order $\mathcal{O}(s^2)$ 
in the second cumulant and the first term of 
$\langle\!\langle g^3 \rangle\!\rangle$ agree with 
Ref.~\onlinecite{macedo94}. For the symmetry class
A (\mbox{$m_0=2$}, \mbox{$m_l=1$}), the subleading term of 
$\langle\!\langle g^3 \rangle\!\rangle$ coincides with 
Ref.~\onlinecite{macedo02}. 

For the BdG classes CI and DIII
(\mbox{$m_0=2,m_l=0$} and \mbox{$m_0=2,m_l=2$}) the coefficients
of the form $a_{k_{1}}^{(m)},a_{k_{1},k_{2}}^{(m)}$ with $m\le -1$
equal zero. By a similar argument as in Appendix B one may show that this 
implies that all the factors of the form $a_{k_{1},\ldots,k_{n}}^{(m)}$ with 
$n\ge 3$ and any $m$ vanish.

Thus in the case of the symmetry classes CI and DIII, the cumulants higher 
than the second do not contain components which are analytic in $s$ at $s=0$. 
For these symmetry classes, the three lowest cumulants contain a contribution 
that is nonanalytic at $s=0$. \cite{macedo02} We consider it likely 
that there exists a residual effect of this nonanalytic behavior in the 
higher-order cumulants too so that these cumulants do not vanish, 
but we cannot prove this by our perturbative method. If such a residual 
effect exists it is special to the cumulants higher than the second in the 
context of classes CI and DIII that the nonperturbative components are 
actually the leading contributions in the diffusive regime. This absence of 
an analytical expression would indicate that the diagrammatic or semiclassical 
pictures \cite{altland96} do not lend themselves to the interpretation of 
these higher-order spectral correlations in the BdG wires with TR symmetry.

The factor $(m_0-2)/m_0$ is a consequence of W(A)L.
For the WD ensemble with broken TR symmetry (\mbox{$m_0=2$}, \mbox{$m_l=1$})
we obtain only even (odd) powers of $s$ in $\langle\!\langle g^n\rangle\!\rangle$ 
with $n$ even (odd), in agreement with Refs.~\onlinecite{gopar95}
and \onlinecite{macedo94} whereas for the symmetry classes AI, AII, C, and D 
all the powers are present.
\subsection{Current cumulants}
As a second application of Eq.~(\ref{eq:recua}) we have calculated the 
current cumulants $C_j$ familiar from the context of full counting statistics
and defined \cite{belzig02}  in terms of the generating function 
$S(\chi)=-\ln[\sum_{N}P_{t_{0}}(N)\exp(iN\chi)]$, 
\begin{eqnarray}
C_{j}&\equiv&-(-i)^{j}\frac{\partial^j}{\partial\chi^j}S(\chi)\bigg{|}_{\chi=0}
.
\end{eqnarray}
Here $P_{t_{0}}(N)$ is the probability of $N$ electrons traversing through the 
sample in a time $t_{0}$ while $\chi$ is a so-called counting field. Since we 
have
\begin{equation}
\left\langle\sum_{i}\tau_i^j\right\rangle_s=\langle T_{j}\rangle_s
=-\sum_{m=-\infty}^{1} a_j^{(m)}s^{-m},
\end{equation}
the equation \cite{lee95}
\begin{eqnarray}
C_{j}&=&\frac{deVt_0}{h}\left\langle\sum_{i}\left[\tau(1-\tau)\frac{d}{d\tau}
\right]^{j-1}\tau\bigg{|}_{\tau=\tau_i}\right\rangle_s
\label{leeeq}
\end{eqnarray}
yields for the current cumulants
\begin{equation}
\begin{split}
C_{1}=&Q_0+c_0+c_1,\\
C_{2}=&\frac{1}{3}Q_0+\frac{1}{15}c_{0}-\frac{3}{7}c_1, \  
C_{3} = \frac{1}{15}Q_0+\frac{1}{315}c_0+\frac{23}{105}c_1,\\
C_{4}=&-\frac{1}{105}Q_0-\frac{11}{1575}c_0-\frac{401}{3465}c_1,\\
C_{5}=&-\frac{1}{105}Q_0-\frac{1}{1485}c_0+\frac{18\ 101}{315\ 315}c_1,\\
C_{6}=&\frac{1}{231}Q_0+\frac{47\ 221}{14\ 189\ 175}c_0-\frac{24\ 433}{945\
  945}c_1,\\
C_{7}=&\frac{27}{5005}Q_0+\frac{811}{2\ 027\ 025}c_0+\frac{62\ 993}{4\ 922\
  775}c_1,\\
C_{8}=&-\frac{3}{715}Q_0-\frac{1\ 790\ 851}{516\ 891\ 375}c_0-\frac{664\
  157}{70\ 509\ 285}c_1,\\
C_{9}=&-\frac{233}{36\ 465}Q_0-\frac{98\ 299\ 813}{206\ 239\ 658\
  625}c_0-\frac{1\ 095\ 799}{204\ 105\ 825}c_1,\\
C_{10}=&\frac{6823}{969\ 969}Q_0+\frac{23\ 610\ 799\ 591}{3\ 781\ 060\ 408\
  125}c_0\\
&+\frac{10\ 053\ 185\ 861}{3\ 478\ 575\ 575\ 475}c_1.
\end{split}
\end{equation}
The next correction terms are of the order $\mathcal{O}(s^2)$. Here we have 
adopted the notations
\begin{eqnarray}
Q_0&=&\frac{I_0t_0}{e},\
I_0
=\frac{e^2}{h}\frac{d}{s}V,\
c_0=\frac{(m_0-2m_l)deVt_0}{3m_0 h},\nonumber\\
c_1&=&\frac{[(3m_0-8m_l)(m_0-2)+4m_l(m_l-2)]deVt_0s}{45m_0^{2}h}.\nonumber\\
\end{eqnarray}
Equations ~(\ref{eq:recua}) and (\ref{leeeq}) provide a straightforward 
derivation for the leading contributions of $C_j$ since for these terms,
in Eq.~(\ref{eq:recua}), only the sums in square brackets 
with \mbox{$n=m=j=n_1=p_1=1$} contribute.
For the leading contributions of $C_j$, our
results agree with those in Ref.~\onlinecite{lee95}.
 It can be shown by Eq.~(\ref{eq:recua}) that the 
correction terms of certain order $\mathcal{O}(s^{-m})$ 
depend similarly on $m_0$ and $m_l$ with all the $j$'s (through 
$c_0, c_1,\ldots$)
and differ only by the numerical coefficients. This was proved 
in Ref.~\onlinecite{imamura02} for the terms of the order 
$\mathcal{O}(s^0)$. While the weak-localization corrections to the current 
cumulants have been studied before, e.g., in Ref.~\onlinecite{lerner88}, 
to our knowledge, the exact numerical values for $j>2$ have not been reported.
Except for the first two cumulants, the ratio of the numerical factors before 
the universal correction term $c_0$ and before the bare first cumulant $Q_0$ 
is for even cumulants of the order, but smaller than unity and for odd 
cumulants smaller by about a factor of ten.
\section{Discussion}
In conclusion, we have presented a recursion equation covering the asymptotic
behavior of the higher-order mesoscopic fluctuations 
for seven universality classes. We evaluated the 
values of the conductance cumulants $\langle\!\langle g^n\rangle\!\rangle$ with 
$n\le 6$ and the weak-localization corrections to the current 
cumulants $C_j$ for $j\le 10$.
We discovered two qualitative results: (1) conductace cumulants of order 
larger than two and smaller than $1/s$ with $s$ the length of the wire in 
units of the number of channels times mean free path scale with $s$ with one 
less power than expected on the basis of naive scaling analysis and (2) the 
same
cumulants are all vanishing in an expansion in powers of $s$ in the two BdG
symmetry classes characterized by TR symmetry, the fact from which we 
conjecture pure nonanalytic dependence on $s$.
As far as the noninteracting DMPK model is valid, $P(g)$ deviates from the 
Gaussian shape only slightly.
These deviations may, however, in principle, be detected by generating a
large number of uncorrelated disorder realizations by repeatedly heating 
and cooling the sample. \cite{mailly92} 
\begin{acknowledgments}
M.P.V.S. acknowledges the financial support of Magnus Ehrnrooth Foundation and 
the Foundation of Technology (TES, Finland). We thank the Center for
Scientific Computing for computing resources.
\end{acknowledgments}
\appendix
\section{Derivation of the recursion eq. (9)}
The DMPK equation (\ref{eq:dmpk}) may be expressed through $\tau_i$s
instead of $\lambda_i$s (for the WD ensembles, see
Ref.~\onlinecite{mello91}). Rewriting Eq.~(A1) of 
Ref.~\onlinecite{imamura01}, which is a direct consequence of the DMPK
equation, in terms of $\tau_i$s yields for 
\mbox{$\exp(-\mbox{\boldmath{$\mathrm{q}$}}
\cdot\mbox{\boldmath{$\mathrm{T}$}})$} an 
evolution equation
\begin{widetext}
\begin{eqnarray}
&&\frac{(m_0 N+1+m_l-m_0)}{2N}\partial_s \langle 
\exp(-\mbox{\boldmath{$\mathrm{q}$}}\cdot
\mbox{\boldmath{$\mathrm{T}$}})\rangle_s^{(m_0,m_l)}=
\bigg{\langle}\bigg\{\sum_{k=1}^{\infty}kq_k \frac{m_0}{2}
\bigg[\bigg(\sum_{l=0}^{k-1}T_{k-l}T_{l+1}\bigg)-
\bigg(\sum_{l=0}^{k-2}T_{k-l-1}T_{l+1}\bigg)
\bigg]+\nonumber\\
&&\sum_{k,l=1}^{\infty}klq_kq_l(T_{k+l}-T_{k+l+1})
+\sum_{k=1}^{\infty}k q_k \bigg[\bigg(1-\frac{m_0}{2}\bigg)k(T_{k+1}-T_{k})+
\frac{(1-m_l)}{2}T_{k+1}+\frac{(2m_l-m_0)}{2}T_k\bigg]\bigg\}
\exp(-\mbox{\boldmath{$\mathrm{q}$}}\cdot\mbox{\boldmath{$\mathrm{T}$}})
\bigg{\rangle}_s^{(m_0,m_l)}.\nonumber\\
\end{eqnarray}
\end{widetext}
For the CGF $\varphi_{s}(\mbox{\boldmath{$\mathrm{q}$}})$, the terms of the 
order $\mathcal{O}(N^{m+1})$ imply
\begin{widetext}
\begin{eqnarray}
\label{eq:dsvarphi}
&&\frac{m_0}{2N}\partial_{s}\varphi_{s}^{(m)}+\frac{(1+m_l-m_0)}{2N}
\partial_{s}\varphi_{s}^{(m+1)}=\sum_{k_1=1}^{\infty}k_1q_{k_{1}}
\bigg\{\frac{m_0}{2}\bigg[
\sum_{l=0}^{k_1-1}\bigg(\sum_{j=m}^{1}
\frac{\partial\varphi_{s}^{(j)}}{\partial q_{k_1-l}}
\frac{\partial\varphi_{s}^{(m-j+1)}}{\partial q_{l+1}}+
\frac{\partial^2\varphi_{s}^{(m+1)}}{\partial q_{k_1-l}\partial
  q_{l+1}}\bigg)\nonumber\\
&&+\sum_{k_2=1}^{\infty}k_2q_{k_2}
\frac{\partial\varphi_{s}^{(m+1)}}{\partial q_{k_1+k_2+1}}
\bigg]-\frac{m_0}{2}[k_1\rightarrow k_1-1]
+\frac{(m_0-2)}{2}k_1\bigg
(\frac{\partial\varphi_s^{(m+1)}}{\partial q_{k+1}}-\frac{\partial\varphi_s^{(m+1)}}{\partial q_{k}}\bigg)
+\frac{(m_l-1)}{2}\frac{\partial\varphi_{s}^{(m+1)}}{\partial q_{k_1+1}}
\nonumber\\  
&&+\frac{(m_0-2m_l)}{2}
\frac{\partial\varphi_{s}^{(m+1)}}{\partial q_{k_1}}\bigg\}.
\end{eqnarray}
\end{widetext}
Making use of the expression for 
$\varphi_{s}^{(m)}(\mbox{\boldmath{$\mathrm{q}$}})$ given by Eqs.~(\ref{nexp}) 
and (\ref{eq:finalexp}) and considering the terms of the order
$\mathcal{O}(s^{-m}\prod_{i=1}^{n}q_{k_i})$ we arrive, after a lengthy algebra,
at an algebraic recursive equation, our Eq.~(\ref{eq:recua}).
\section{Cancellation of the terms $\mathcal{O}(s^{n-2})$ for 
$\langle\!\langle g^n\rangle\!\rangle$ with $n\ge 3$}
Let us consider Eq.~(\ref{eq:recua}) with a factor of the form 
$a_{q_1,q_2,\ldots,q_n}^{(2-n)}$ on the left-hand side (lhs). On the 
right-hand side (rhs) of Eq.~(\ref{eq:recua}), only the sums in square 
brackets and the $r$ sum on 
the third line may then contribute. There are no factors of the form 
$a_{q_1,q_2,\ldots,q_n}^{(3-n)}$ in Eq.~({\ref{eq:finalexp}}). The factors 
$a_{q_1,q_2,q_3}^{(-1)}$ vanish for all the WD classes. \cite{macedo94} 
Equation (\ref{eq:recua}) implies \mbox{$a_{q_1,q_2,q_3}^{(-1)}=0$} for all 
the symmetry classes since the terms containing $m_l$ cannot contribute to 
$a_{q_1,q_2,q_3}^{(-1)}$. It is thus sufficient to show that on the rhs, 
the joint contribution of (i) sums containing products of factors of the 
form $a_{q_1}^{(1)}$ and $a_{q_1,q_2,\ldots,q_n}^{(2-n)}$, (ii) sums 
containing products of factors $a_{q_2,q_3}^{(0)}$ and 
$a_{q_1,q_2,\ldots,q_{n-1}}^{(3-n)}$, and (iii) factors of the form 
$a_{q_1,q_2,\ldots,q_{n-1}}^{(3-n)}$ vanishes. 

For the induction argument, let us consider Eq.~(\ref{eq:recua})
with \mbox{$m=2-n_0$}, \mbox{$n=n_0$} fixed and assume that we have
\mbox{$a_{q_1,q_2,\ldots,q_{n_0-1}}^{(3-n_0)}=0$} for all positive integer 
values of $q_i$. 
The induction assumption implies that
the contributions of the forms (ii) and (iii) vanish such that 
we have \mbox{$a_{q_1,q_2,\ldots,q_{n_0}}^{(2-n_0)}=0$} with
\mbox{$q_1=q_2=\cdots =q_{n_0}=1$}. This implies that also the contribution (i)
vanishes and we have \mbox{$a_{q_1,q_2,\ldots,q_{n_0}}^{(2-n_0)}=0$}
with arbitrary $q_i$s.
The induction assumption holds with $n_0=4$ since we 
have $a_{q_1,q_2,q_3}^{(-1)}=0$. Thus we obtain
$a_{q_1,q_2,\ldots,q_n}^{(2-n)}=0$ for arbitrary $n\ge 3$.

\end{document}